\documentclass[11pt,a4j]{article}
\usepackage[dvips]{graphicx}
\usepackage{mathrsfs}
\usepackage{eucal}
\usepackage{amsmath,amsthm,amssymb}
\usepackage{wrapfig}
\usepackage[dvips]{color}
\usepackage{cite}
\usepackage{framed}
\usepackage{tabularx}
\usepackage[sectionbib]{chapterbib}
\usepackage{threeparttable}
\usepackage{float}
\definecolor{shadecolor}{gray}{0.80}
\DeclareMathVersion{chem}
\SetSymbolFont{letters}{chem}{OT1}{cmr}{m}{n}

\newcommand{\CF}{C_{\textit{\textsf{\hspace{-0.3mm}F}}}}

\begin{document}

\renewcommand{\figurename}{\small{Fig.}~}
\renewcommand{\thefootnote}{$\dagger$\arabic{footnote}}

\begin{flushright}
\textit{Estimation of Expansion Factor in Melt State}
\end{flushright}
\vspace{1mm}

\begin{center}
\setlength{\baselineskip}{25pt}{\LARGE\textbf{Excluded Volume Effects of Branched Molecules}}
\end{center}
\vspace*{-7mm}
\begin{center}
\setlength{\baselineskip}{18pt}{\Large\textbf{Estimation of Expansion Factor in Melt State}}
\end{center}

\vspace*{0mm}
\begin{center}
\large{Kazumi Suematsu} \vspace*{2mm}\\
\normalsize{\setlength{\baselineskip}{12pt} 
Institute of Mathematical Science\\
Ohkadai 2-31-9, Yokkaichi, Mie 512-1216, JAPAN\\
E-Mail: suematsu@m3.cty-net.ne.jp,  Tel/Fax: +81 (0) 593 26 8052}\\[8mm]
\end{center}

\hrule
\vspace{0mm}
\begin{flushleft}
\textbf{\large Abstract}
\end{flushleft}
The expansion factor, $\alpha^{2}=\langle s_{N}^{2}\rangle/\langle s_{N}^{2}\rangle_{0}$, of branched molecules in the melt state is estimated. The equilibrium expansion factor is determined as the point in which all the inhomogeneity terms of the osmotic potential, $\Delta G_{osmotic}$, go to zero. Numerical analysis shows that $\log\,\alpha=0.082\,\log N+\text{const.}$ for $10^{3}\le N\le 10^{7}$, giving $\alpha\cong \text{const.}\,N^{1/12}$ so that $\langle s_{N}^{2}\rangle^{1/2}\propto N^{1/3}$ which coincides with the value for the critical packing density.\\[-3mm]
\begin{flushleft}
\textbf{\textbf{Key Words}}:
\normalsize{Branched Molecules/ Excluded Volume Effects/ Osmotic Potential/ Inhomogeneity Term/ Equilibrium Expansion Factor}\\[3mm]
\end{flushleft}
\hrule
\vspace{3mm}
\setlength{\baselineskip}{13pt}
\section{Introduction}
This paper deals with the volume expansion in concentrated solutions. The expansion factor of branched molecules is estimated from the new point of view, making use of the the basic equation of the chemical potential.

Our knowledge on polymer solutions is that (i) the excluded volume effects arise from the wild inhomogeneity of segment concentration, and (ii) the magnitude of the effects is controlled by the solvent-solute interaction\cite{Flory} (the enthalpy parameter $\chi$).

\section{Theoretical}
Let $V_{1}$ denote the volume of a solvent molecule and $V$ the space volume. The basic thermodynamic formula is
\begin{equation}
\Delta G_{osmotic}=\frac{kT}{V_{1}}\int\left\{-\left(1-\chi\right)\mathscr{J}_{1}+\left(1/2-\chi\right)\mathscr{J}_{2}+\frac{1}{6}\mathscr{J}_{3}+\cdots\right\}\delta V\label{BEV-Basic1}
\end{equation}
where $\mathscr{J}_{k}'s\, (=v_{hill}^{k}-v_{valley}^{k})$ represent inhomogeneity terms with $v$ denoting the local volume fraction of segments and the subscripts $hill$ and $valley$ the more concentrated and the more dilute region respectively.
The molecular dimensions are determined by the force balance between the osmotic potential, $\Delta G_{osmotic}$, and the elastic potential, $\Delta G_{elastic}$. However the osmotic potential is more fundamental than the elastic potential, because under $\Delta G_{osmotic}=0$, no volume expansion is possible.

Since monomers are joined by chemical bonds, dilute polymer solutions are necessarily accompanied by the wild inhomogeneity of the segment concentration, while such inhomogeneity is never allowed in the non-solvent state because of the physical instability. So in the vicinity of the melt state, all $\mathscr{J}_{k}'s$ should converge to 0. Then when we plot $\mathscr{J}$ as a function of the expansion factor ($\alpha$), the equilibrium state that gives $\alpha_{eq}$ should occur in the close proximity of $\mathscr{J}=0$. This $\alpha_{eq}$ is just the quantity which we are going to seek.

 By definition, $v_{2}=V_{2}\rho_{N}$, where the subscript 2 denotes the polymer and $\rho_{N}$ is the segment density at the coordinate $(x, y, z)$ around a molecule ($N$ being the number of segments). Within the framework of the Gaussian approximation\cite{Ishihara, Debye}, $\rho_{N}$ has the form:
\begin{equation}
\rho_{N}=N\left(\frac{\beta}{\pi\alpha^{2}}\right)^{3/2}\sum_{\{a, b, c\}}\exp\left\{-\frac{\beta}{\alpha^{2}}\left[(x-a)^{2}+(y-b)^{2}+(z-c)^{2}\right]\right\}\label{BEV-Basic2}
\end{equation}
The constants $\{a, b, c\}$ represent the location of the center of gravity of individual polymer molecules in the system, and $\beta=3/2\langle s_{N}^{2}\rangle_{0}$ with $\langle s_{N}^{2}\rangle_{0}$ being the radius of gyration of an unperturbed molecule; $\langle s_{N}^{2}\rangle_{0}=\frac{1}{6}\CF\, Nl^{2}$ for linear chains ($\CF$ being the characteristic constant), but $\langle s_{N}^{2}\rangle_{0}\cong (\frac{(f-1)\pi}{2^{3}(f-2)})^{1/2} N^{\frac{1}{2}}l^{2}$ ($N\rightarrow$ large) for branched molecules\cite{Zim, Dobson, Kajiwara}.

According to the preceding work\cite{Kazumi1}, we solve the present problem using the lattice model. Polymer molecules are arranged on the sites of the simple cubic lattice. Then the maxima and the minima of the segment concentration are located on the $x=y=z$ line\cite{Kazumi1}. Applying eq. (\ref{BEV-Basic2}) to eq. (\ref{BEV-Basic1}), we can evaluate the density fluctuation as a function of $x$ and $\alpha$. For the present purpose we use a hypothetical model-polymer-system: a branched polymer that is made from the cyclotrimerization of bisphenol A dicyanate\cite{Stutz}, and $N-$methylpyrrolidone as the solvent . The employed parameters are listed in Table \ref{BPADC-Table} (the mean bond length, $\bar{l}$, and the enthalpy parameter, $\chi$, are arbitrary). In this simulation, we identify the segment with the repeating unit.

\begin{center}
  \begin{threeparttable}[h]
    \caption{Parameters of a hypothetical branched polymer solution ($d=3$)}\label{BPADC-Table}
  \begin{tabular}{l l c r}
\hline\\[-1.5mm]
& \hspace{10mm}parameters & notations & values \,\,\,\,\\[2mm]
\hline\\[-1.5mm]
branched polymer & volume of a solvent (NMP\tnote{\,a}\,\,\,) & $V_{1}$ & \hspace{5mm}160 \text{\AA}$^{3}$\\[1.5mm]
& volume of a segment & $V_{2}$ & \hspace{5mm}387 \text{\AA}$^{3}$\\[1.5mm]
& mean bond length & $\bar{l}$ & \hspace{5mm}10 \text{\AA}\,\,\,\\[1.5mm]
& enthalpy parameter & $\chi$ & \hspace{5mm}0 \,\,\,\,\,\,\,\\[2mm]
\hline\\[-6mm]
   \end{tabular}
    \vspace*{2mm}
   \begin{tablenotes}
     \item a. N-methylpyrrolidone.
   \end{tablenotes}
  \end{threeparttable}
  \vspace*{4mm}
\end{center}

\begin{figure}[H]
\begin{center}
\includegraphics[width=17cm]{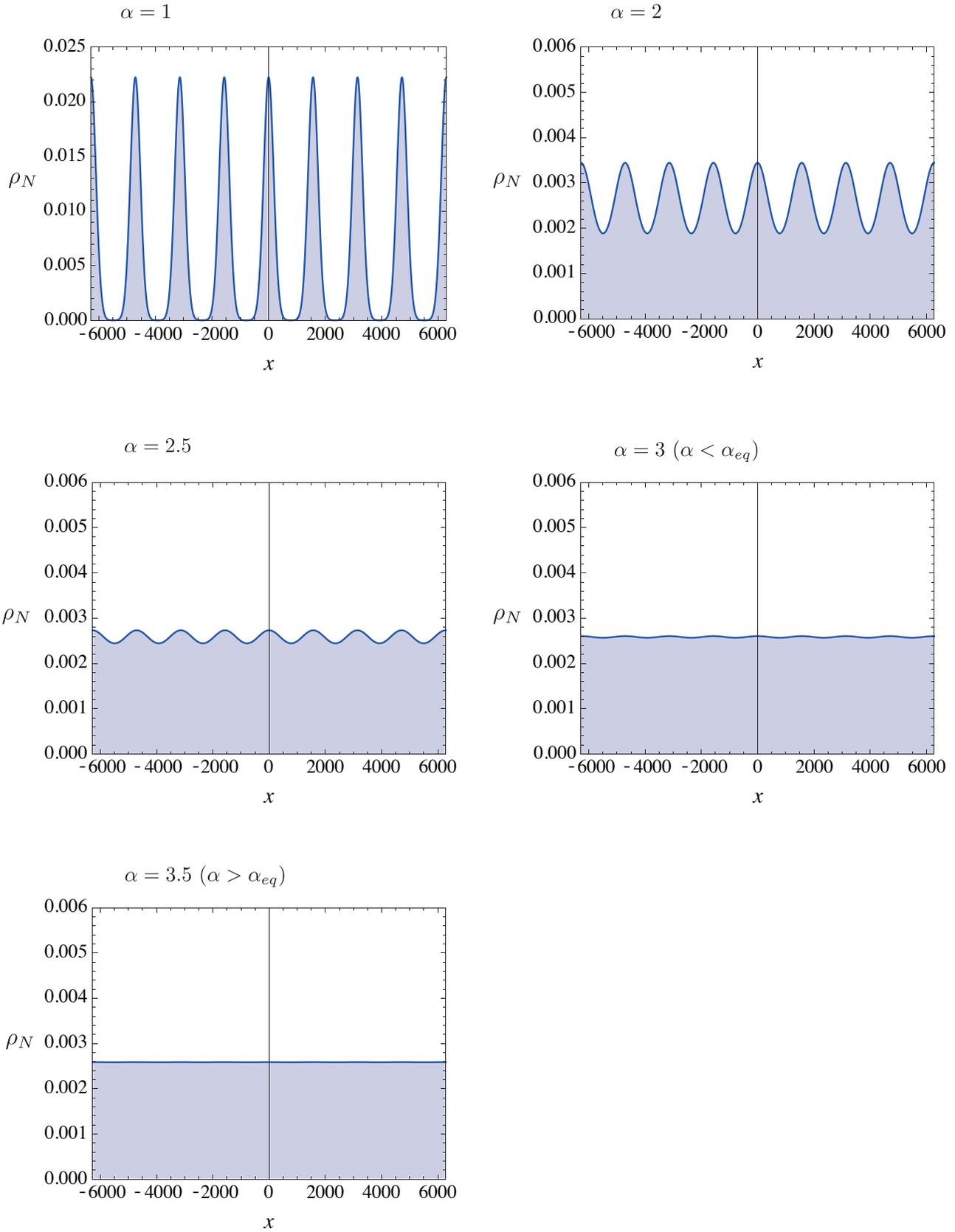}
\caption{The variation of the inhomogeneity in the melt state as a function of $\alpha$ for the hypothetical branched polymer with $N=10^{7}$. The equilibrium point is estimated as $\alpha\cong 3.1$}\label{DisappearanceInhomogeneity}
\end{center}
\end{figure}

Examples of the density inhomogeneity in the melt state as a function of $\alpha$ are illustrated in Fig. \ref{DisappearanceInhomogeneity} for the polymer having $N=10^{7}$. To determine the equilibrium point, we have calculated $\Delta\rho/\bar{\rho}$ where $\Delta\rho=\rho_{hill}-\rho_{valley}$ with $\rho$ denoting the local density of segments around a polymer molecule and $\bar{\rho}$ the average density in the whole system. The mean separation, $p$, on the lattice between the centers of gravity of molecules can be calculated by the equation $\bar{\phi}=V_{2}N/p^{3}$, where $\bar{\phi}$ is the average volume fraction of segments in the system. In the present work, we confine ourselves to $\bar{\phi}\rightarrow 1$, and the equilibrium point is taken so that $\Delta\rho/\bar{\rho}<0.01$. The rate of the change of the required quantity, $\Delta\rho/\bar{\rho}$, as against the variation of $\alpha$ is so rapid that there is no difficulty in locating the equilibrium point. In the example of Fig. \ref{DisappearanceInhomogeneity}, $\alpha_{eq}$ is in the vicinity of $\alpha=3.1$.

\begin{figure}[H]
\begin{center}
\includegraphics[width=9cm]{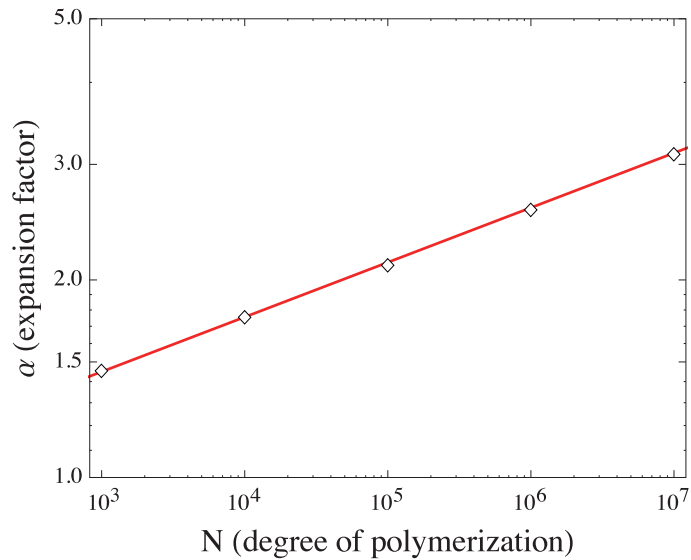}
\caption{The variation of the equilibrium expansion factor in the melt state as a function of $N$ for the hypothetical branched polymer. The solid line obeys the linear equation, $\log\,\alpha=0.082\,\log N+\text{const.}$, giving $\alpha\cong \text{const.} N^{\frac{1}{12}}$.}\label{AlphainMelt}
\end{center}
\end{figure}

The numerical results for $\alpha_{eq}$ thus estimated are plotted in Fig. \ref{AlphainMelt} for the interval, $10^{3}\le N\le 10^{7}$. Excellent linearity is found according to the equation:
\begin{equation}
\log\,\alpha=0.082\,\log N+\text{const.}\label{BEV-Basic3}
\end{equation}
From this result we can calculate the exponent $\nu$ defined by $\langle s_{N}^{2}\rangle^{1/2}\propto N^{\nu}$\cite{Stauffer, Redner, deGennes, Issacson, Seitz, Parisi, Daoud, Lubensky}: The observed value, $0.082$, is close to $1/12$, giving $\alpha\cong \text{const.} N^{\frac{1}{12}}$, so that since $\langle s_{N}^{2}\rangle_{0}^{1/2}\propto N^{1/4}\,\, (N\rightarrow\infty)$\cite{Zim, Dobson, Kajiwara} for randomly branched polymers, we have
\begin{equation}
\langle s_{N}^{2}\rangle^{1/2}=\alpha\,\langle s_{N}^{2}\rangle_{0}^{1/2}\propto N^{1/3}\hspace{5mm}(\text{for}\,\, N\rightarrow\infty)\label{BEV-Basic4}
\end{equation}

It is of course possible to assert that $\alpha$ may reach the true equilibrium well below the value calculated above because of the presence of the elastic potential, $\Delta G_{elastic}$, so that the wild inhomogeneity is still alive against the physical instability. For this objection it might be more proper for us to state that the exponent is less than 1/3, i.e., $\nu\le 1/3$. Quite on the other hand, the accommodation problem requires $\nu\ge 1/3$, or else the packing density diverges as $\rho_{N}\propto N^{1-\nu\hspace{0.2mm}d}$ for $N\rightarrow\infty$ ($d$ being the space dimensions). So we may conclude that the exponent can be equated exactly with the value of the critical packing density\cite{Kazumi3}:
\begin{equation}
\nu=\frac{1}{3}\label{BEV-Basic5}
\end{equation}


\end{document}